# A note on classification of proper homothetic vector fields in Kantowski-Sachs and Bianchi type III Lorentzian manifolds

Ghulam Shabbir and Fouzia Iqbal

Faculty of Engineering Sciences, GIK Institute of Engineering Sciences and Technology, Topi, Swabi, Khyber Pakhtunkhwa, Pakistan. Email: shabbir@giki.edu.pk

**Abstract**

A complete study of Kantowski-Sachs and Bianchi type III space-times according to their proper homothetic vector fields is given by using direct integration technique. Using the above mentioned technique we have shown that very special classes of the above space-times admit proper homothetic vector fields. The dimension of homothetic vector fields is five.

## 1. INTRODUCTION

The aim of this paper is to explore all the possiblilties when the Kantowski-Sachs and Bianchi type III space-times admit proper homothetic vector fields. Homothetic vector fields which presevers metric structure up to the constant conformal factor carries significant information and plays an important role in Einstein's theory of relativity. It is important to study homothetic symmetry. Direct integration techinque is used to study proper homothetic vector fields in the Kantowski-Sachs and Bianchi type III space-times.

Throughout $M$ represents a four dimensional, connected, Hausdorff space-time manifold with Lorentz metric $g$ of signature (-, +, +, +). The curvature tensor associated with $g_{ab}$, through the Levi-Civita connection, is denoted in component form by $R^a{}_{bcd}$, and the Ricci tensor components are $R_{ab} = R^c{}_{acb}$. The usual covariant, partial and Lie derivatives are denoted by a semicolon, a comma and the symbol $L$, respectively. Round and square brackets denote the usual symmetrization and skew-symmetrization, respectively.

Any vector field $X$ on $M$ can be decomposed as

$$X_{a;b} = \frac{1}{2}h_{ab} + F_{ab} \qquad (1)$$

where $h_{ab}(=h_{ba}) = L_X g_{ab}$ and $F_{ab}(=-F_{ba})$ are symmetric and skew symmetric tensors on $M$, respectively. If

$$h_{ab} = 2C g_{ab}, \quad C \in R$$

equivalently,

$$g_{ab,c} X^c + g_{cb} X^c_{,a} + g_{ac} X^c_{,b} = 2C g_{ab} \tag{2}$$

then $X$ is called a homothetic vector field on $M$. If $X$ is homothetic and $C \neq 0$ then it is called proper homothetic while $C = 0$ it is Killing [1]. Further consequences and geometrical interpretations of (2) are explored in [1]. It also follows from (2) that [1]

$$L_X R^a{}_{bcd} = 0, \qquad L_X R_{ab} = 0.$$

## 2. MAIN RESULTS

Consider the space-times in the usual coordinate system $(t, r, \theta, \phi)$ with line element [2]

$$ds^2 = -dt^2 + A(t)dr^2 + B(t)\left[d\theta^2 + f^2(\theta) d\phi^2\right], \tag{3}$$

where $A$ and $B$ are no where zero functions of $t$ only. For $f(\theta) = \sin\theta$ or $f(\theta) = \sinh\theta$ the above space-time (3) become Kantowski-Sachs or Bianchi type III space-times, respectively. The above space-time admits four independent Killing fields which are [4, 5]

$$\frac{\partial}{\partial r}, \frac{\partial}{\partial \phi}, \cos\phi \frac{\partial}{\partial \theta} - \frac{f'}{f}\sin\phi \frac{\partial}{\partial \phi}, \sin\phi \frac{\partial}{\partial \theta} + \frac{f'}{f}\cos\phi \frac{\partial}{\partial \phi}, \tag{4}$$

where prime denotes the derivative with respect to $\theta$. A vector field $X$ is said to be a homothetic vector field if it satisfies equation (2). One can write (2) explicitly and using (3) to get

$$X^0{}_{,0} = C, \tag{5}$$

$$X^0{}_{,1} - A X^1{}_{,0} = 0, \tag{6}$$

$$X^0{}_{,2} - B X^2{}_{,0} = 0, \tag{7}$$

$$X^0{}_{,3} - B f^2(\theta) X^3{}_{,0} = 0, \tag{8}$$

$$\dot{A} X^0 + 2 A X^1{}_{,1} = 2 C A, \tag{9}$$

$$A X^1{}_{,2} + B X^2{}_{,1} = 0, \tag{10}$$

$$A X^1{}_{,3} + B f^2(\theta) X^3{}_{,1} = 0, \tag{11}$$

$$\dot{B} X^0 + 2 B X^2{}_{,2} = 2 C B, \tag{12}$$

$$X^2{}_{,3} + f^2(\theta) X^3{}_{,2} = 0, \tag{13}$$

$$\dot{B} X^0 + f'(\theta) X^2 + X^3{}_{,3} = 2 C B f(\theta), \tag{14}$$

where dot denotes the derivative with respect to $t$. Equations (5), (6), (7) and (8) give

$$\left.\begin{array}{l} X^0 = Ct + N^1(r,\theta,\phi), \quad X^1 = N^1{}_r(r,\theta,\phi) \int \dfrac{1}{A(t)} dt + N^2(r,\theta,\phi), \\[6pt] X^2 = N^1{}_\theta(r,\theta,\phi) \int \dfrac{1}{B(t)} dt + N^3(r,\theta,\phi), \\[6pt] X^3 = \dfrac{N^1{}_\phi(t,\theta,\phi)}{f^2(\theta)} \int \dfrac{1}{B(t)} dt + N^4(r,\theta,\phi), \end{array}\right\} \tag{15}$$

where $N^1(r,\theta,\phi), N^2(r,\theta,\phi), N^3(r,\theta,\phi)$ and $N^4(r,\theta,\phi)$ are functions of integration. In order to determine $N^1(r,\theta,\phi), N^2(r,\theta,\phi), N^3(r,\theta,\phi)$ and $N^4(r,\theta,\phi)$ we need to integrate the remaining six equations. To avoid lengthy calculations here we will only present results for full details see [3]

**Case (1)** Five independent homothetic vector fields:

In this case the space-times (3) take the form

$$ds^2 = -dt^2 + \alpha(Ct + c_5)^{\frac{2(C-c_6)}{C}} dr^2 + \beta(Ct + c_5)^2 \left[d\theta^2 + f^2(\theta) d\phi^2\right], \tag{16}$$

and homothetic vector fields in this case are

$$\begin{array}{l} X^0 = Ct + c_5, \quad X^1 = c_6 r + c_1, \quad X^2 = c_4 \cos\phi + c_3 \sin\phi, \\[6pt] X^3 = -c_4 \dfrac{f'}{f} \sin\phi + c_3 \dfrac{f'}{f} \cos\phi + c_2, \end{array} \tag{17}$$

where $c_1, c_2, c_3, c_4, c_5, c_6, \alpha, \beta \in \Re(\alpha, \beta \neq 0)$. The above space-time (16) admits five independent homothetic vector fields in which four are Killing vector fields and one is proper homothetic vector field which is $t\dfrac{\partial}{\partial t}$. In this case one can easily see that proper homothetic vector field exist for a very special of $A(t)$ and $B(t)$ which are given in equation (16).

**Case (2)** Five independent homothetic vector fields:

In this case the space-times (3) take the form

$$ds^2 = -dt^2 + \alpha \, dr^2 + \beta t^2 \left[d\theta^2 + f^2(\theta) d\phi^2\right], \tag{18}$$

and homothetic vector fields in this case are

$$X^0 = Ct, \quad X^1 = Cr + c_1, \quad X^2 = c_4 \cos\phi + c_3 \sin\phi,$$
$$X^3 = -c_4 \frac{f'}{f} \sin\phi + c_3 \frac{f'}{f} \cos\phi + c_2, \tag{19}$$

where $c_1, c_2, c_3, c_4, \alpha, \beta \in \Re (\alpha, \beta \neq 0)$. The above space-time (18) admits five independent homothetic vector fields in which four are Killing vector fields and one is proper homothetic vector field which is $t\frac{\partial}{\partial t} + r\frac{\partial}{\partial r}$. In this case one can easily see that proper homothetic vector field exist for a very special of $A(t)$ and $B(t)$.

**Case (3)** Five independent homothetic vector fields:

In this case the space-times (3) take the form

$$ds^2 = -dt^2 + \alpha(Ct + c_5)^2 dr^2 + \beta(Ct + c_5)^2 \left[d\theta^2 + f^2(\theta) d\phi^2\right], \tag{20}$$

and homothetic vector fields in this case are

$$X^0 = Ct + c_5, \quad X^1 = c_1, \quad X^2 = c_4 \cos\phi + c_3 \sin\phi,$$
$$X^3 = -c_4 \frac{f'}{f} \sin\phi + c_3 \frac{f'}{f} \cos\phi + c_2, \tag{21}$$

where $c_1, c_2, c_3, c_4, c_5, \alpha, \beta \in \Re(\alpha, \beta \neq 0)$. The above space-time (20) admits five independent homothetic vector fields in which four are Killing vector fields and one is proper homothetic vector field which is $t\frac{\partial}{\partial t}$. In this case one can easily see that proper homothetic vector field exist for a very special of $A(t)$ and $B(t)$, given in equation (20).

## Summary

In this paper a study of Kantowski-Sachs and Bianchi type III space-times according to their proper homothetic vector fields is given by using the direct integration technique. From the above study we obtain that Kantowski-Sachs and Bianchi type III space-times admit five independent homothetic vector fields which are given in equations (17), (19) and (21) (see for details cases (1), (2) and (3)).